\documentclass{aastex63}
\usepackage{multirow}
\usepackage{threeparttable}

\newcommand{\angstrom}{{$\rm \mathring A$}}

\received{X XX, XXXX}
\revised{X XX, XXXX}
\accepted{X XX, XXXX}
\submitjournal{PASP}

\shorttitle{The Compact Star-Forming Galaxies at $2<z<3$ in 3D-HST/CANDELS}
\shortauthors{Lu et al.}

\begin{document}

\title{The Compact Star-Forming Galaxies at $2<z<3$ in 3D-HST/CANDELS: AGN and Non-AGN Physical Properties}

\correspondingauthor{Qirong Yuan and Guanwen Fang}
\email{yuanqirong@njnu.edu.cn, wen@mail.ustc.edu.cn}

\author{Shiying Lu}
\affil{\rm School of Physics Science and Technology, Nanjing Normal University, Nanjing  210023, China; yuanqirong@njnu.edu.cn}

\author{Yizhou Gu}
\affil{\rm School of Physics Science and Technology, Nanjing Normal University, Nanjing  210023, China; yuanqirong@njnu.edu.cn}

\author{Guanwen Fang}
\altaffiliation{Guanwen Fang and Shiying Lu contributed equally to this work}
\affil{\rm Institute for Astronomy and History of Science and Technology, Dali University, Dali 671003, China; wen@mail.ustc.edu.cn}

\author{Qirong Yuan}
\affil{\rm School of Physics Science and Technology, Nanjing Normal University, Nanjing  210023, China; yuanqirong@njnu.edu.cn}

\author{Min Bao}
\affil{\rm School of Physics Science and Technology, Nanjing Normal University, Nanjing  210023, China; yuanqirong@njnu.edu.cn}

\author{Xiaotong Guo}
\affil{\rm School of Astronomy and Space Science, Nanjing University, Jiangsu 210093, China}

\begin{abstract}
We investigate the differences in the stellar population properties, the structure, and the environment between massive compact star-forming galaxies (cSFGs) with or without active galactic nucleus (AGN) at $2<z<3$ in the five 3D-HST/CANDELS fields.
In a sample of 221 massive cSFGs, we constitute the most complete AGN census so far, identifying 66 AGNs by the X-ray detection, the mid-infrared color criterion, and/or the SED fitting, while the rest (155) are non-AGNs.
Further dividing these cSFGs into two redshift bins, i.e., $2<z<2.5$ and $2.5 \leq z<3$, we find that in each redshift bin the cSFGs with AGNs have similar distributions of the stellar mass, the specific star formation rate, and the ratio of $L_{\rm IR}$ to $L_{\rm UV}$ to those without AGNs.
After having performed a two-dimensional surface brightness modeling for those cSFGs with X-ray-detected AGNs (37) to correct for the influence of the central point-like X-ray AGN on measuring the structural parameters of its host galaxy, we find that in each redshift bin the cSFGs with AGNs have comparable distributions of all concerned structural parameters, i.e., the Sersic index, the 20\%-light radius, the Gini coefficient, and the concentration index, to those without AGNs. With a gradual consumption of available gas and dust, the structure of cSFGs, indicated by the above structural parameters, 
seem to be slightly more concentrated with decreasing redshift. At $2<z<3$, the similar environment between cSFGs with and without AGNs suggests that their AGN activities are potentially triggered by internal secular processes, such as gravitational instabilities or/and dynamical friction.
\end{abstract}


\section{Introduction}
\label{Sect1}

Recently, massive star-forming galaxies (SFGs) with compact structure\footnote{The compactness of a galaxy is usually defined as $\log_{\rm 10} (M_{\ast}/(r_{\rm e} \times \sqrt{q})^{1.5}/[M_{\odot}\;\rm kpc^{-1.5}])>10.45$, or $r_{\rm e}< 2$ kpc, etc, where $M_*$ is the galactic stellar mass, $r_{\rm e}$ the effective radius, and $q$ the axis ratio of the given galaxy.}
at $z>2$ have been suggested to be the direct progenitors of compact quiescent galaxies (cQGs) at $z=1.5$-$3$ (e.g., \citealt{Whitaker+12, Barro+13,Barro+14,van+der+Wel+14,Fang+15,van+Dokkum+15,Lu+19}).
The compact SFG (cSFG) candidates have centrally concentrated light profiles and spheroidal morphologies similar to cQGs (\citealt{Barro+14, Barro+17}).
Several theories have been proposed for achieving high stellar densities observed in cQGs, including gas-rich interactions and disk instabilities (\citealt{Tremonti+07,Heckman+11, Barro+13,Barro+14,Wellons+15,Zolotov+15}). The prevalence of active galactic nuclei (AGNs) in cSFGs suggests that AGN activity might have played an important role in quenching star formation, possibly by feedbacks such as outflows (e.g., \citealt{Kocevski+17,Hopkins+06,Rangel+14}).

Being proposed by \citet{Barro+13}, gas-rich cSFGs at $z\sim 2$-$3$ are formed by violent dissipation processes (e.g., mergers or disk-instabilities) that induce building a compact starburst bulge. The subsequent highly efficient star formation and/or AGN feedback can rapidly quench cSFGs, and fade them into cQGs.
An important question is that what possible effects the AGN has exerted on the evolution of cSFG. It is found that the X-ray-selected AGNs at $z\geq 2$ favor more spheroid-dominated hosts with high levels of dust obscuration (\citealt{Grogin+05,Pierce+07,Kocevski+12,Rangel+14,Kocevski+17}). The intense AGN activities can halt the star formation in cSFGs through powerful radiation and outflow, which is called quasar-mode feedback \citep[e.g.,][]{Hopkins+06,Cai+13}. Along with this process, the supermassive black hole grows (e.g., \citealt{Rangel+14,Kocevski+17}) and the star formation and structure of cSFGs can be regulated (\citealt{Barro+13,Barro+14,Fang+15,Chang+17,Habouzit+19}). However, results from \citet{Mahoro+17} and \citet{Mahoro+19} suggest that the X-ray-selected AGNs with far-infrared emission have been found to reside in the SFGs with higher star formation rates (SFRs) than inactive galaxies in the green valley, which is inconsistent with previous works (e.g., \citealt{Ellison+16,Gu+18}).
Furthermore, \citet{Diamond-Stanic+12} discover a cSFG at $z\sim 0.6$ with outflow of $\geq$ 1000 $\rm km$ $\rm s^{-1}$, and suggest that the stellar feedback associated with compact starburst, in the form of radiation pressure from massive stars and ram pressure by stellar winds, is sufficient to produce the high-velocity outflow without the need of AGN feedback.

Using a single selection method or criterion may miss a part of the AGN census in the universe.
In most previous works, AGNs are widely selected by their X-ray emissions \citep{Kocevski+12,Barro+13,Barro+14,Rangel+14,Fang+15,Kocevski+17},
while due to the presence of large amounts of dust, which absorbs the UV/optical radiation, obscured AGNs can be selected in infrared \citep{Houck+05,Higdon+05,Yan+05,Bornancini+17}.
A large fraction of AGNs is found in highly luminous and obscured galaxies, especially at higher redshift where major merger is prevalent \citep{Springel+05,Hopkins+06,Gilli+07,Rangel+14}.
The AGN fraction missed by the UV/optical selections varies considerably, ranging from 15\% to over 50\% \citep{Richards+03, Brown+06, Glikman+04,Glikman+07}. As such, the heavily obscured AGNs missed in X-ray might be identified and supplemented by other techniques, such as the mid-infrared (MIR) color selection and the composite spectral energy distribution (SED) fitting.

Many MIR-based selection criteria are designed to target heavily obscured active galaxies.
The MIR emission, relatively insensitive to the intervening obscuration, can trace the reprocessed radiation of dust heated by AGN \citep{Lacy+04,Lacy+07,Stern+05, Donley+07, Donley+08, Donley+12}. The MIR selection therefore can recover a substantial fraction of luminous unobscured and obscured AGNs missed in the X-ray surveys.
\citet{Lacy+04} propose a MIR color-color diagram and define a region populated by optically selected quasars from the SDSS. On the other hand,
\citet{Stern+05} find a clear vertical spur populated by bright active galaxies in the [3.6]-[4.5] versus [5.8]-[8.0] color-color magnitude diagram.
While \citet{Donley+12} find that the more luminous the AGNs, the redder their MIR colors, the precise location of a source in the MIR color space depends on the relative AGN/host contributions, its redshift, the reddening of the host and AGN components, and the host galaxy type.
The moderate- to high-redshift SFGs are inadequately separated from AGNs in both \citet{Lacy+04} and \citet{Stern+05} MIR color-color spaces.
Considering a more secure MIR power-law selection ($f_{\nu} \propto \nu^{\alpha}$, $\alpha \leq -0.5$, e.g., \citealt{Donley+07, Donley+08, Park+10}), \citet{Donley+12} revise the color-color selection by attaching a cut power-law box to improve the completeness and reliability.
Gradually, the MIR selection as a potentially powerful technique is accepted by many studies to identify obscured AGNs (e.g., \citealt{Fang+15,Bornancini+17,Chang+17}).
Moreover, the SED fitting technique using the observed multi-band SED is now a widely-used technique to decompose the AGN/host components and derive properties of AGN host galaxies (\citealt{Ciesla+15,Malek+18,Gao+19}). \cite{Guo+20} derive the physical parameters of 791 X-ray sources (518 AGNs and 273 normal galaxies) by fitting their SEDs using the Code Investigating GALaxy Evolution code (CIGALE 0.12.1, \citealt{Burgarella+05, Noll+09, Boquien+19}). Six AGN candidates are selected from the 273 normal galaxies based on their SEDs, and two of the six AGN candidates can also be identified by their X-ray variabilities \citep{Ding+18}.

Although the AGN fraction of cSFGs has been investigated by various previous works, there remains large uncertainties on the physical properties of cSFGs with and without AGNs at $2 < z < 3$. In order to thoroughly understand the difference of the physical properties of cSFGs with AGNs and non-AGNs, and to disentangle the AGN effect on their hosts, in this work we perform a statistical analysis on cSFGs with AGNs and non-AGNs selected from the five deep fields of the 3D-HST/CANDELS program \citep{Grogin+11, Koekemoer+11, Skelton+14,Momcheva+16}.
For a sample of 221 massive ($M_* \geq 10^{10} M_{\odot}$) cSFGs at $2 < z < 3$, constructed by the rest-frame UVJ diagram \citep{Williams+09} and compactness criterion in \citet{Barro+14}, we select a large sample of 66 AGNs in these cSFGs by criteria of the X-ray detection, the MIR color \citep{Donley+12} and/or the SED fitting using CIGALE.
Distributions of the stellar population, the structural parameters, and the environment for cSFGs with AGNs and non-AGNs, are analysed in two redshift intervals with $\Delta z = 0.5$.

This paper is laid out as follows.
In Section~\ref{Sect2}, the 3D-HST/CANDELS observation and data are introduced. In Section~\ref{Sect3}, the sample selection is described, including the compact galaxy selection in subsection~\ref{Sect3.1} and the AGN selection in subsection~\ref{Sect3.2}.
Distributions of various physical properties for cSFGs with AGNs and non-AGNs are compared in Section~\ref{Sect4}, including the stellar population properties in subsection~\ref{Sect4.1}, the parametric structural measurements in subsection~\ref{Sect4.2},
and the non-parametric structural measurements in subsection~\ref{Sect4.3}. In Section~\ref{Sect5}, we analyse the environmental effect on the cSFGs with and without AGNs.
Discussion on the physical properties of cSFGs is presented in Section~\ref{Sect6} and summary in Section~\ref{Sect7}.
Throughout this paper, we assume a flat $\Lambda$CDM cosmology with $\rm \Omega_{M} = 0.3$, $\rm \Omega_{\Lambda} = 0.7$, and $H_{0} = 70\, \rm km\;s^{-1}~Mpc^{-1} $. All magnitudes given in this paper are in AB system.

\section{Observations and Data}
\label{Sect2}
The 3D-HST and CANDELS Multi-Cycle Treasury programs \citep{Grogin+11,Koekemoer+11} target five premier existing survey fields (i.e., AEGIS, COSMOS, GOODS-N, GOODS-S, and UDS) of $\sim$ 900 arcmin$^2$ via deep imaging of more than 250,000 galaxies with WFC3/IR and ACS. The 3D-HST program provides a large amount of data sets, including photometries \citep{Skelton+14} and grism spectra \citep{Momcheva+16}. Some value-added data products are also available, such as the stellar population and structural parameters (\citealt{van+der+Wel+12,Whitaker+14}).

The multi-band databases in the 3D-HST/CANDELS have already been provided. The data can also be retrieved from Rainbow database \citep{Barro+11}, a central repository of CANDELS-related data that can be accessed via a web-based interface\footnote{http://rainbowx.fis.ucm.es}.
The photometries are heterogeneous as the exact combination of bandpasses varies from field to field. The multiwavelength photometry includes broadband data from the UV (from CFHT, KPNO, VLT, and WFI2.2m), optical (from CFHT, HST, Keck, Subaru, VISTA, UKIRT, and WFI2.2m), near-to-mid IR (from CFHT, HST, Spitzer, Subaru, and VISTA), and far-IR (from Spitzer and Herschel) observations. In each field, there are approximately 20-45 band photometric data, spanning from UV to FIR, which are enough to perform SED fitting in this work. The multi-wavelength filters used in the SED fitting for each field are tabulated in Table~\ref{tab1}.

For the MIR photometries used to pick out AGN candidate in this work, the observed Infrared Array Camera (IRAC) images in different fields, which are from different surveys, are coordinated by \citet{Skelton+14}. The IRAC images at 3.6 and 4.5 $\mu m$ within the five fields are furnished with the deep infrared Spitzer Extended Deep Survey (\citealt{Ashby+13}), which covers a total area of 1.46 $\rm deg^2$ to a depth of 26 mag.
The IRAC images at 5.8 and 8.0 $\mu m$ for the Extended Groth Strip survey, the COSMOS Spitzer survey, the GOODS Spitzer survey, and the Spitzer Public Legacy Survey of UKIDSS UDS are provided by \citet{Barmby+08}, \citet{Sanders+07}, \citet{Dickinson+03}, and \citet{Skelton+14}, respectively.
All IRAC fluxes integrated by \citet{Skelton+14} have taken the contamination of nearby sources into consideration.

\begin{table}[!t]
\centering
\caption{The multi-wavelength filters used for the five 3D-HST/CANDELS fields}\label{tab1}
\begin{threeparttable}
\begin{tabular}{clll}
\hline\hline
\textbf{Field} & \textbf{Telescope} & \textbf{Instrument} & \textbf{Filters}\\
\hline
\multirow{4}{*}{AEGIS} & CFHT & MegaCam/MIRcam & $u, g, i, r, z, J, H, K_s$\\
&HST     & ACS/WFC3& F606W, F814W, F140W, F125W, F160W\\
&Spitzer & IRAC/MIPS    & 3.6, 4.5, 5.8, 8.0, 24, 70 $\mu m$ \\
\hline
\multirow{8}{*}{COSMOS} & CFHT & MegaCam/MIRcam & $u, g, i, r, z, J, H, K_{s}$\\
\cline{4-4}
&Subaru &Suprime-Cam & $B, V, r', i', z'$ \\
& & & IA427, IA464, IA484, IA505, IA527, IA574 \\
& & & IA624, IA606, IA709, IA738, IA767, IA827 \\
\cline{4-4}
&HST & ACS/WFC3 & F606W, F814W, F140W, F120W, F160W \\
&VISTA&VISTA & $Y, J, H, K_s$  \\
&Spitzer & IRAC/MIPS    & 3.6, 4.5, 5.8, 8.0, 24, 70 $\mu m$ \\
&Herschel & PACS/SPIRE & 100, 160, 250, 350, 500 $\mu m$\\
\hline
\multirow{7}{*}{GOODS-N} & KPNO4m &KPNO4m &U\\
&Keck & LRIS & $G, R_s$ \\
&Subaru & Suprime-Cam/MOIRCS & $B, V, R_c, I_c, z'$ \\
\cline{4-4}
&HST  & ACS/WFC3& F435W, F606W, F775W, F850LP, F140W\\
& & &F125W, F160W\\
\cline{4-4}
&Spitzer & IRAC/MIPS    & 3.6, 4.5, 5.8, 8.0, 24, 70 $\mu m$ \\
&Herschel & PACS/SPIRE & 100, 160, 250, 350, 500 $\mu m$\\
\hline
\multirow{9}{*}{GOODS-S} &VLT& VIMOS/ISAAC & $U, R, J, H, K_s$\\
& WFI2.2m & WFI2.2m & $U38, B, V, R_c, I$\\
& CFHT & WIRCam & $J, K_s$\\
\cline{4-4}
& Subaru & Suprime-Cam & IA427, IA445, IA505, IA527, IA550, IA574 \\
& & & IA624, IA651, IA679, IA738, IA767, IA797\\
\cline{4-4}
&HST & ACS/WFC3 & F435W, F606W$^a$, F775W, F850LP$^a$ \\
& & & F814W, F140W, F125W, F160W \\
&Spitzer & IRAC/MIPS    & 3.6, 4.5, 5.8, 8.0, 24, 70 $\mu m$ \\
&Herschel & PACS/SPIRE & 70, 100, 160, 250, 350, 500 $\mu m$\\
\hline
\multirow{6}{*}{UDS} & CFHT & MegaCam & U \\
& Subaru & Suprime-Cam & $B, V, R_c, i', z'$\\
& HST & ACS/WFC3 & F606W, F814W, F140W, F125W, F160W \\
& UKIRT & WFCAM & $J, H, K_s$\\
&Spitzer & IRAC/MIPS    & 3.6, 4.5, 5.8, 8.0, 24, 70 $\mu m$ \\
&Herschel & PACS/SPIRE & 100, 160, 250, 350, 500 $\mu m$\\
\hline
\end{tabular}
 \begin{tablenotes}
        \footnotesize
        \item[$^a$] There are two kinds of filters from different surveys. One is for GOODS from \cite{Giavalisco+04} and the other is for CANDELS from \cite{Grogin+11} and \cite{Koekemoer+11} (also see \citealt{Skelton+14} for more details).
 \end{tablenotes}
\end{threeparttable}
\end{table}

The derived data products are also provided by the 3D-HST/CANDELS program.
The `best' redshift (\texttt{z\_best}) catalog mergers the \citet{Momcheva+16} grism-based redshift with the \citet{Skelton+14} photometric one. We prefer to take the spectroscopic redshift (\texttt{z\_spec}) or the grism redshift (\texttt{z\_max\_grism}) if available, otherwise we use the photometric redshift (\texttt{z\_phot}).
Compared with the average error of photometric redshift, $\Delta z /(1+z) \approx 0.02$, \citet{Momcheva+16} derive the \texttt{z\_max\_grism} with high accuracy, i.e., $\Delta z /(1+z) \approx 0.003$, using a modified version of the \textrm{EAZY} templates \citep{Brammer+08}.
Given redshift, the rest-frame colors can be derived with the \textrm{EAZY} templates as well. The stellar masses and other parameters of the stellar population are estimated with the \textrm{FAST} code \citep{Kriek+09} on the basis of several assumptions, concerning exponentially declining star formation histories, the \citet{Calzetti+00} dust attenuation and the \citet{Bruzual+Charlot+03} stellar population synthesis models with the \citet{Chabrier+03} initial mass function (\textrm{IMF}) and solar metallicity. Additionally, the structural parameters of a galaxy, such as the S\'{e}rsic index ($n$), the effective radii ($r_{\rm e}$), and the axis ratio ($q\equiv b/a$), are inferred from the 3D-HST/CANDELS WFC3 \textit{H}-band image by \citet{van+der+Wel+12} using GALFIT (\citealt{Peng+02}).

\section{Sample Selection}
\label{Sect3}
\subsection{Compact Galaxy Selection}\label{Sect3.1}
Based on the multi-wavelength data in the five 3D-HST/CANDELS fields, we firstly select 1764 massive ($M_\ast \geq 10^{10}~M_{\odot}$) galaxies at $2<z<3$ with good photometric quality (i.e., $\texttt{use\_phot~=~1}$), good morphological fit (i.e., GALFIT flag = 0 or 1), and low contamination from neighbors in the IRAC bands (i.e., $\texttt{contam\_flag~=~0}$) to ensure high sample completeness and robust structural measurements. The completeness above the mass threshold, $M_* \geq 10^{10}~M_{\odot}$, is $\sim 90 \%$ up to the considered highest redshift \citep{Grogin+11,Wuyts+11,Newman+12,Barro+13,Pandya+17}.

\begin{figure*}[!t]
  \begin{center}
  \includegraphics[width=0.95\textwidth, angle=0 ]{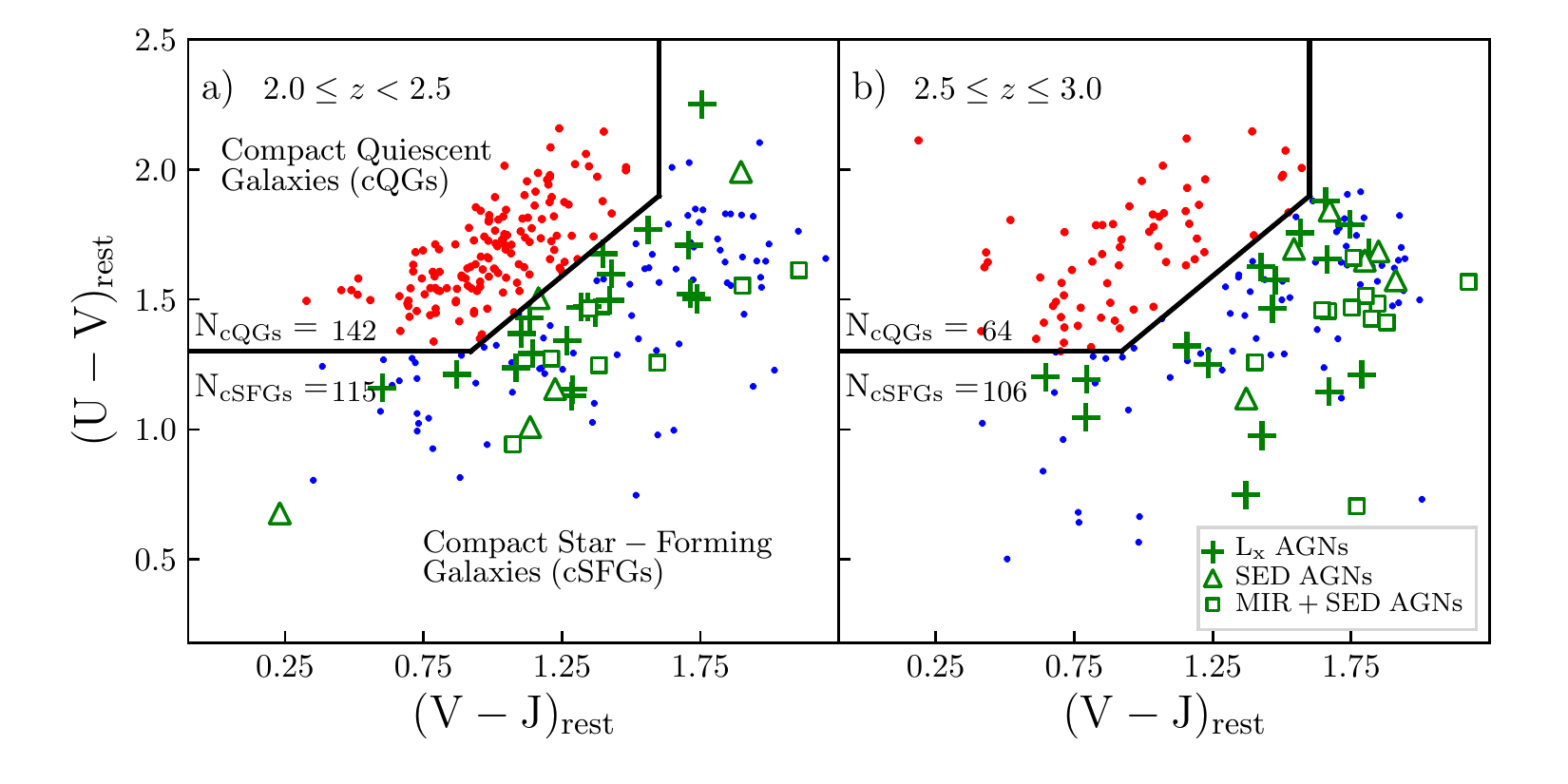}
  \end{center}
  \caption{The rest-frame UVJ diagram for compact massive galaxies in two redshift bins. The criteria (solid black lines) from \citet{Williams+09} are used to separate cSFGs from cQGs (red dots). The numbers of cQGs and cSFGs are shown near the horizontal boundaries between them. The cSFGs with non-AGNs are marked by blue dots, while those with AGNs are shown as green markers, including X-ray-selected AGNs (crosses), MIR+SED selected AGNs (squares) identified using the MIR criteria plus the SED fitting by \citealt{Donley+12}, and SED AGNs (triangles) selected using the SED fitting only (see subsection~\ref{Sect3.2} for detail).}
  \label{fig1}
\end{figure*}

To investigate the physical properties of cSFGs, we first utilize the \citet{Williams+09} rest-frame UVJ diagram to select SFGs. Many previous works have suggested that the UVJ diagram can be employed to separate SFGs from QGs, even up to high redshift $z \sim 3$ \citep{Wuyts+07,Williams+09,Whitaker+11,Whitaker+12,Muzzin+13,van+der+Wel+14,Huertas-Company+15}.
Various compactness criteria have been addressed in recent literature (e.g., \citealt{Barro+13,Carollo+13,Quilis+Trujillo+13,Barro+14,van+Dokkum+15,Lu+19}). Throughout the paper, the compactness of a galaxy is quantified as $\Sigma_{1.5} \equiv \log_{\rm 10} (M_{\ast}/(r_{\rm e} \times \sqrt{q})^{1.5}/[M_{\odot}\;\rm kpc^{-1.5}])$, where $r_{\rm e}$ and $q$ are estimated by the GALFIT code (\citealt{van+der+Wel+12}). Finally, we select a sample of 221 cSFGs at $2<z<3$ with $\Sigma_{1.5} > 10.45$ \citep{Barro+14}.
In Figure~\ref{fig1}, the compact samples of both cQGs and cSFGs are exhibited in the rest-frame UVJ diagrams in two redshift bins with an interval of $\Delta z =0.5$.
Only $\sim 4\%$ of cSFGs have spectroscopic redshifts, $\sim 27\%$ have grism redshifts, and the remaining have photometric redshifts.

\subsection{Compact AGN Selection}\label{Sect3.2}
To further analyse the physical properties between cSFGs with and without AGNs, we combine three methods to identify and select as many AGN candidates as possible.

We firstly match the cSFG sample with the X-ray catalogs in 2/7 Ms Chandra Deep Field-North/South survey (\citealt{Xue+16,Luo+17}), 4.6 Ms Chandra COSMOS-Legacy survey (1.8 Ms old C-COSMOS survey plus 2.8 Ms new Chandra ACIS-I observations, \citealt{Civano+16, Marchesi+16}), and 0.8 Ms AEGIS-X Deep survey (\citealt{Nandra+15}). As a result, 37 of 221 cSFGs have X-ray detections with the X-ray luminosity $L_{\rm x} \geq 3\times 10^{42}$ erg $\rm s^{-1}$, which can be identified as X-ray-selected AGNs (hereafter $L_{\rm x}$ AGNs: crosses in Figure~\ref{fig1}). The median and average of $L_{\rm x}$ for these X-ray-detected AGNs are $4.06\times 10^{43}$ erg $\rm s^{-1}$ and $6.93\times 10^{43}$ erg $\rm s^{-1}$, respectively.

Since AGNs prefer to inhabit compact galaxies with higher obscuration at high redshift (e.g., \citealt{Rangel+14}),
the X-ray emission of AGNs could be heavily obscured by dust and gas, while the re-emitted MIR emissions are insensitive to the intervening obscuration.
The IRAC color-color criteria, which can recover most of luminous unobscured and obscured AGNs, have been proposed and adopted by previous works \citep{Lacy+04,Houck+05, Higdon+05, Stern+05, Yan+05, Donley+07, Lacy+07, Donley+08, Donley+12, Bornancini+17, Chang+17}.
Then, we use the IRAC color-color criteria proposed by \citet{Donley+12} and select 40 MIR AGN candidates.  Of the 40 MIR-selected AGNs, 11 have already been identified as $L_{\rm x}$ AGNs.

To further identify potentially omissive or mis-identified AGN candidates selected according to the MIR colors, we directly model the UV to far-IR SEDs of our cSFGs to constrain their AGN contribution, employing the CIGALE code (\citealt{Burgarella+05, Noll+09, Boquien+19}). In this work, following \citet{Guo+20}, for each source in our sample we firstly fit its SED by utilizing the \textit{galaxy} and \textit{galaxy}+\textit{AGN} templates, and then find the best-fit SED iteratively. There are 4 modules in the galaxy templates, which are the star formation history, the \cite{Calzetti+00} dust attenuation, the \cite{Bruzual+Charlot+03} stellar population synthesis models with the \cite{Chabrier+03} IMF and solar metallicity, and the \cite{Dale+14} dust emission. The module for the AGN component is from \cite{Fritz+06}. All above modules are provided by CIGALE and are summarized in Table~\ref{tab2}.

\begin{table}[!t]
\centering
\caption{Modules and relevant parameters in the CIGALE code used for the SED fitting}\label{tab2}
\begin{tabular}{c|l|l|l}
\hline\hline
\textbf{Component} & \textbf{Module} & \textbf{Parameter} & \textbf{Value}\\
\hline
\multirow{14}{*}{\textbf{Galaxy}} & \multirow{5}{*}{sfh (delayed+burst)} & tau\_main (1 Myr)& 20 - 8000 (in steps of 10)\\
 & & tau\_brust (1 Myr) & 10 - 200 (in steps of 1)\\
 & & age\_main (1 Myr) & 200 - 13000 (in steps of 10)\\
 & & age\_brust (1 Myr) & 10 - 200 (in steps of 1)\\
 \cline{3-4}
 & &\multirow{2}{*}{f\_brust (1 Myr)} & 0, 0.0001, 0.0005, 0.001, 0.005, 0.01, \\
  & & &0.05, 0.1, 0.15, 0.20, 0.25, 0.3, 0.40, 0.50\\
 \cline{2-4}
 & \multirow{2}{*}{BC03} & IMF & 1 (Chabrier) \\
 & & metallicity & 0.02 \\
 \cline{2-4}
 &\multirow{2}{*}{dustatt\_calzleit} & \multirow{2}{*}{E\_BV\_nebular (mag)} & 0.005, 0.01, 0.025, 0.05, 0.075, 0.10, 0.15,\\
 & & & 0.20, 0.25, 0.30, 0.35, 0.40, 0.45, 0.50, 0.55, 0.60 \\
 \cline{2-4}
 &\multirow{4}{*}{dl2014} & qpah & 1.12, 1.77, 2.50, 3.19 \\
 & & umin & 5.0, 6.0, 7.0, 8.0, 10.0, 12.0, 15.0, 17.0, 20.0, 25.0\\
 & & alpha& 2.0, 2.1, 2.2, 2.3, 2.4, 2.5, 2.6, 2.7, 2.8 \\
 & & gamma& 0.02 \\
 \hline
 \multirow{9}{*}{\textbf{AGN}} & \multirow{9}{*}{Fritz2006}& r\_ratio &10, 30, 60, 100, 150 \\
 & & tau & 0.1, 0.3, 0.6, 1.0, 2.0, 3.0, 6.0, 10.0\\
 & & beta & -1.00, -0.75, -0.50, -0.25, 0.00\\
 & & gamma & 0.0, 2.0, 4.0, 6.0 \\
 & & opening\_angle & 60, 100, 140 \\
 \cline{3-4}
 & & \multirow{2}{*}{psy} & 0.001, 10.100, 20.100, 30.100, 40.100, \\
 & & & 50.100, 60.100, 70.100, 80.100, 89.990\\
 \cline{3-4}
 & &\multirow{2}{*}{fracAGN} &  0.0, 0.05, 0.1, 0.15, 0.2, 0.25, 0.3, 0.35, 0.4, 0.45, 0.5, \\
 & & & 0.55, 0.6, 0.65, 0.7, 0.75, 0.8, 0.85, 0.9, 0.95, 0.99\\
\hline
\end{tabular}
\end{table}

\begin{figure}[!t]
\begin{minipage}{0.48\linewidth}
  \centerline{\includegraphics[width=1\textwidth]{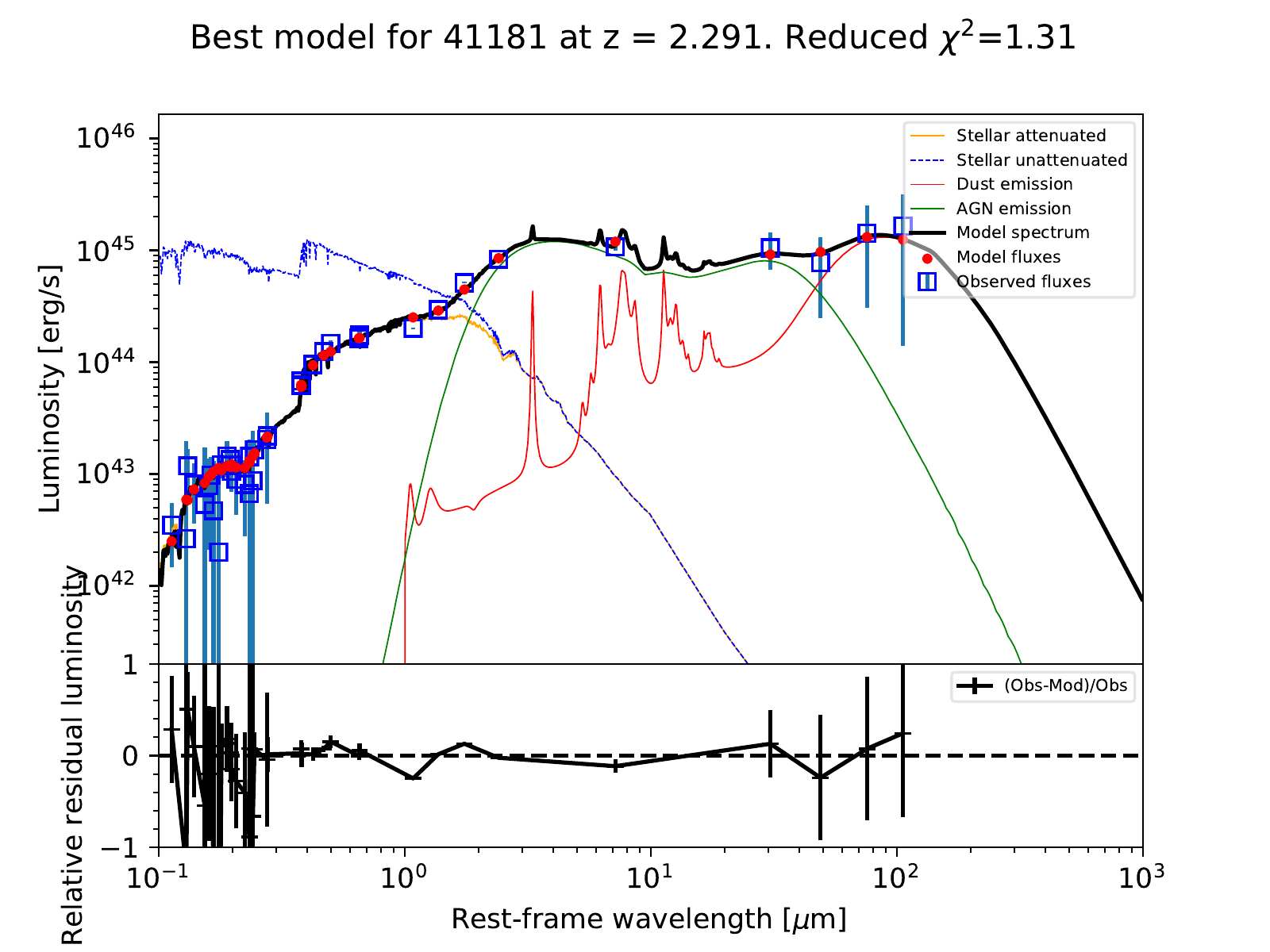}}
  \centerline{(a) A $L_{\rm x}$ AGN (ID 41181) in the GOODS-S field}
\end{minipage}
\hfill
\begin{minipage}{0.48\linewidth}
  \centerline{\includegraphics[width=1\textwidth]{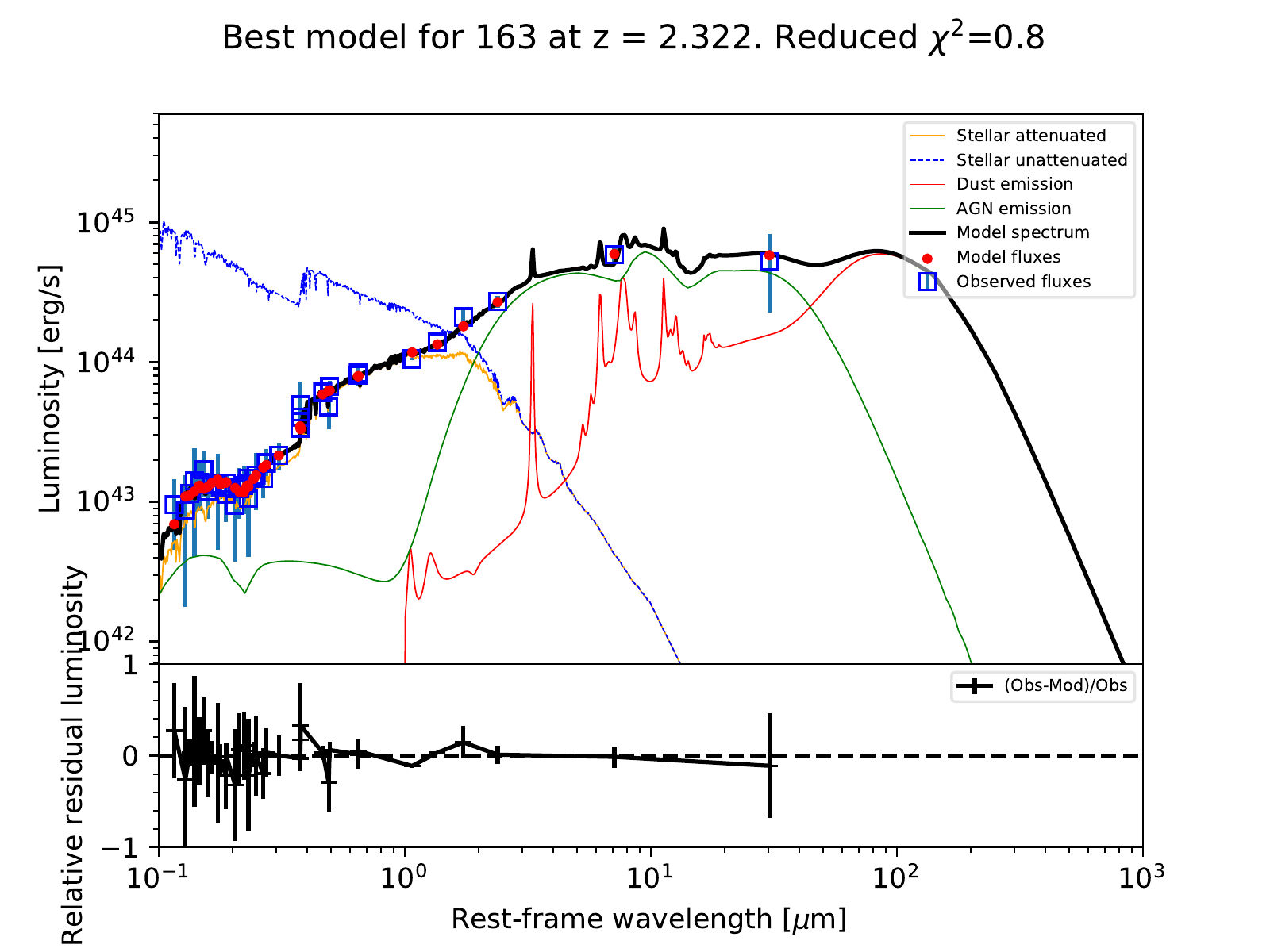}}
  \centerline{(b) A MIR+SED AGN (ID 163) in the COSMOS field}
\end{minipage}
\begin{minipage}{0.48\linewidth}
  \centerline{\includegraphics[width=1\textwidth]{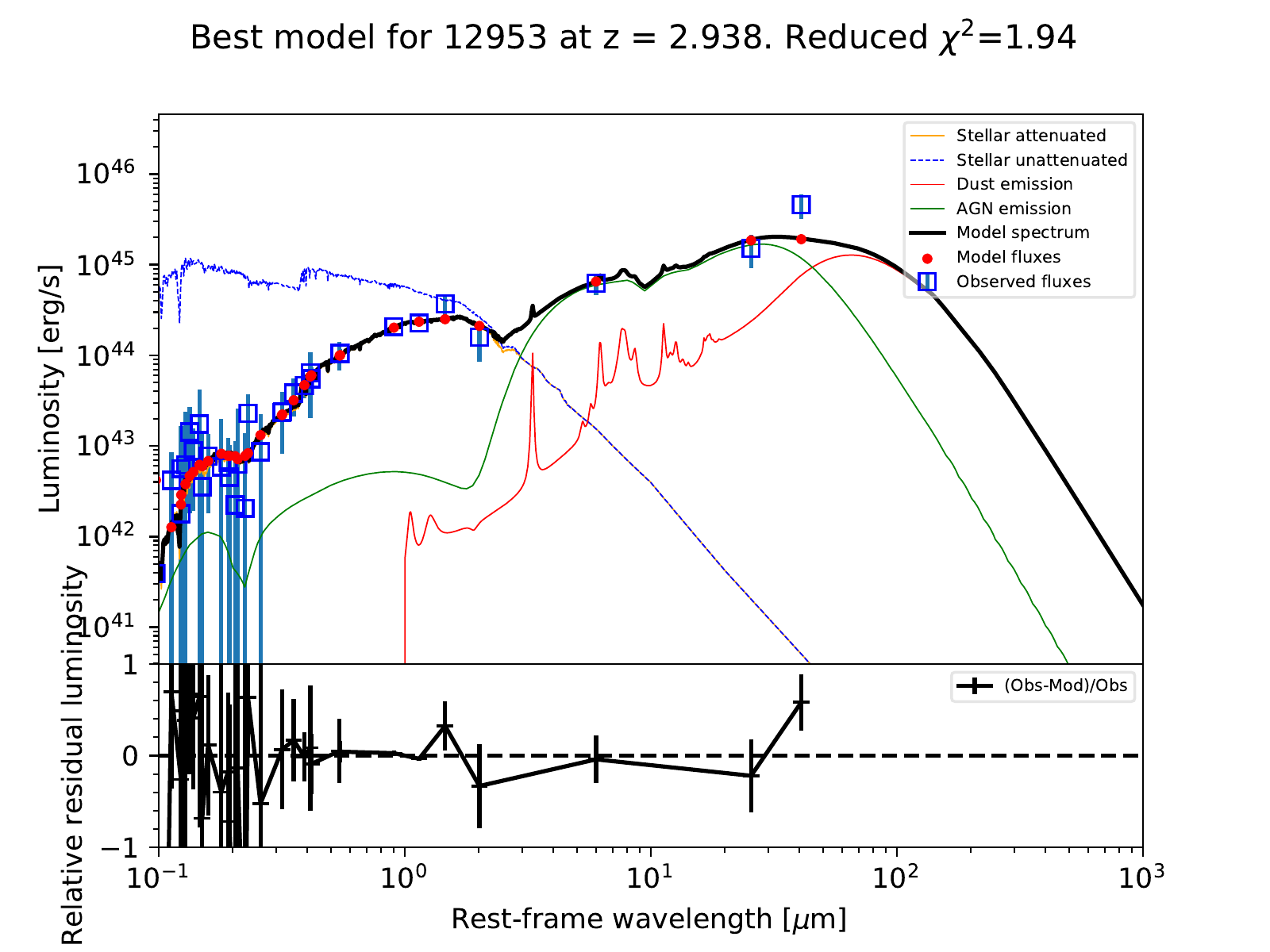}}
  \centerline{(c) ID 12953 with an AGN component}
\end{minipage}
\hfill
\begin{minipage}{0.48\linewidth}
  \centerline{\includegraphics[width=1\textwidth]{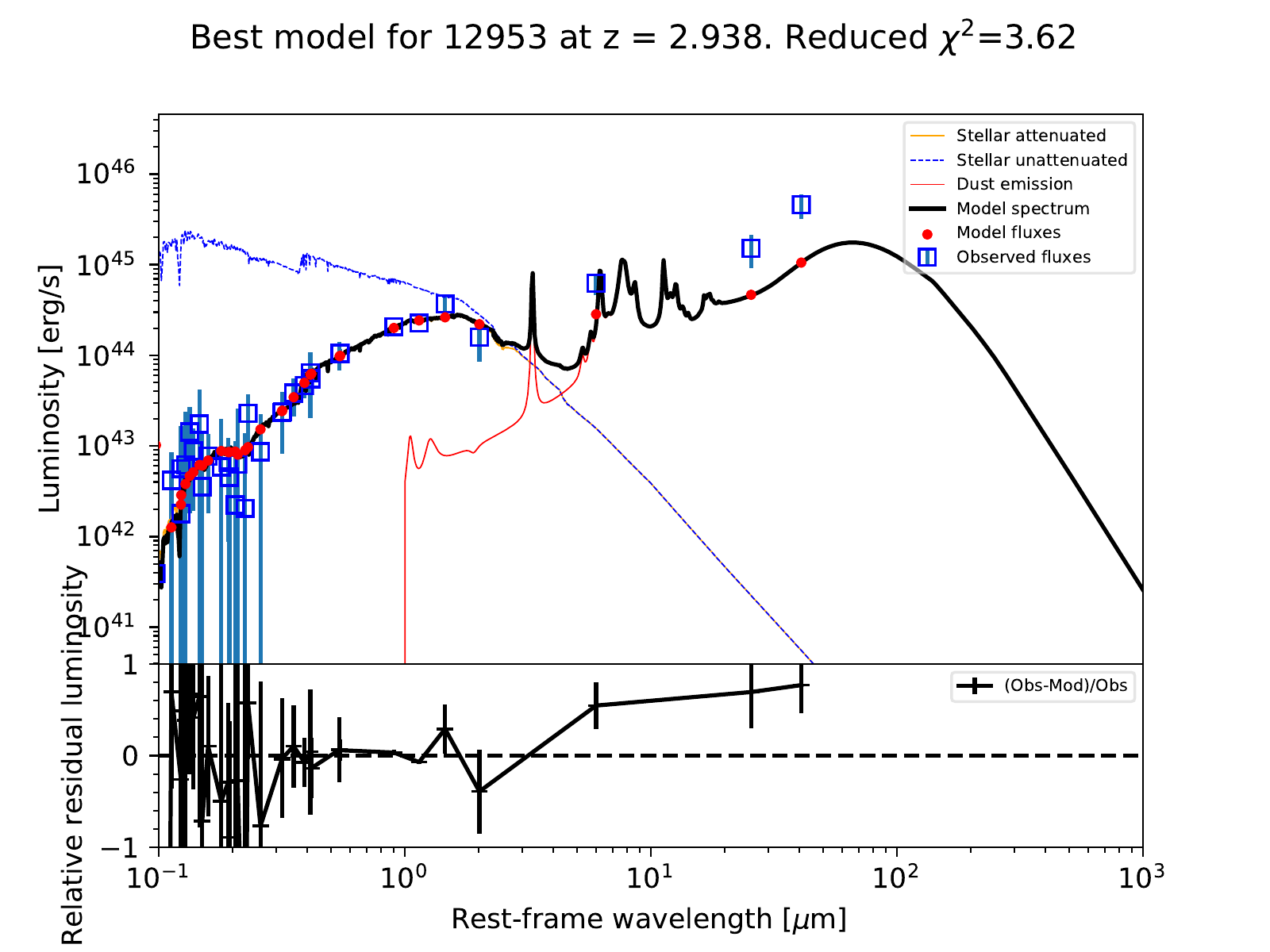}}
  \centerline{(d) ID 12953 without an AGN component }
\end{minipage}
\caption{Examples of the SED fitting for AGN candidates. (a) The source with ID 41181 in the GOODS-S filed can be identified as an AGN by both X-ray and MIR criteria (see subsection~\ref{Sect3.2} for detailed criteria), which is short for ``$L_{\rm x}$ AGN''; (b) The source with ID 163 in the COSMOS field, selected by MIR criteria, may be an AGN candidate and confirmed by the SED fitting, which is short for ``MIR+SED AGN''. The SED fittings with/without AGN component for the source with ID 12953 in the COSMOS field are shown in the panels (c) and (d), respectively. The orange, blue, red, green lines represent attenuated stellar, unattenuated stellar, dust and AGN components, respectively. The black line indicates the best-fit model. The observed and model flux are drawn by blue squares and red dots. The lower panel indicates the residual of the best fit relative to the observed SED.}
\label{fig2}
\end{figure}

For all cSFGs in our sample, we obtain two best-fit SEDs with likelihood of $\chi^2 < 7$ utilizing the \textit{galaxy} and \textit{galaxy}+\textit{AGN} templates, respectively. Then, we seek for potential AGN candidate by comparing the two best-fit SEDs with eyeballing inspection.
If the goodness of fit for a source is not significantly improved when considering the AGN contribution,
it will be classified as a non-AGN. Otherwise, it is classified as an AGN candidate. As a result,
besides the secure 37 X-ray-selected AGNs, 29 AGN candidates are found using the SED fitting. Out of these 29 AGN candidates, 18 of them have been selected using the IRAC colors (i.e., the MIR+SED AGNs, squares in Figure~\ref{fig1}), while 11 are newly identified by the SED fitting (i.e., the SED AGNs, {triangles in Figure~\ref{fig1}}).
Coincidentally, the numbers of the mis-identified and missing AGN candidates selected using the IRAC colors are the same.
Three typical cSFGs with or without an AGN component are illustrated in Figure~\ref{fig2}.
The panel (a) presents the SED of an X-ray-selected AGN with ID 41181 in the GOODS-S field, while that of a MIR+SED AGN with ID 163 in the COSMOS field is shown in panel (b). The best-fit SED results for a SED AGN (ID 12953 in the COSMOS field) with or without an AGN component are compared in the panels (c) and (d) of Figure~\ref{fig2}, respectively. Intuitively, for this SED AGN the fit in the MIR-to-FIR bands is quite bad if there is no AGN component. Therefore, sources analogue to this case are classified as SED AGNs.

Finally, 66 out of 221 cSFGs are selected with AGN component
(see Table~\ref{tab3} for details).
The AGN fraction at $2<z<3$ is $\sim 30 \pm 3.1 \%$, which is higher than that derived solely using a single selection method. In the same redshift interval, \cite{Fang+15} adopt three separate methods, that is the \cite{Stern+05} MIR color-color space, the MIR spectral index, and the X-ray detection ($L_{\rm 0.5-10\; keV}> 10^{41}~\rm erg\;s^{-1}$), to select AGN candidates in the COSMOS field. They find that the corresponding AGN fractions in cSFGs are $27\%$, $24\%$, and $19\%$, respectively, which are indeed lower than our result. In the GOODS-S field, \cite{Barro+14} find that about $47\%$ of cSFGs at $2<z<3$ host an X-ray-detected AGN, while in the same field our AGN fraction is about $56\%$ at $2<z<3$.
Thus it is seemingly more effective on selecting AGN candidates with a combination of multiple methods than a single method.

\begin{table}[!t]
\centering
\caption{The numbers and fraction of the massive cSFGs with AGN and non-AGN at $2<z<3$}\label{tab3}
\begin{tabular}{ccccc}
\hline\hline
redshift & AGN & non-AGN & cSFG & $f_{\rm AGN}$ \\
\hline
$2.0< z < 2.5 $ &32  & 83 &115 &27 $\pm$ 4.2\% \\
$2.5\leq z<3.0$ &34  & 72 &106 &32 $\pm$ 4.5\% \\
\hline
$2.0< z <3.0  $ &66   & 155&221 &30 $\pm$ 3.1\% \\
\hline
\end{tabular}
\end{table}

\section{Physical Properties}
\label{Sect4}
In this section, we analyse the difference of the physical properties of cSFGs with AGNs and non-AGNs at $2 < z < 2.5$ and $2.5 \le z < 3$. The physical properties of galaxies include the stellar mass ($M_*$), the specific star formation rate ($\rm sSFR$), the ratio of $L_{\rm IR}$ to $L_{\rm UV}$ (${\rm IRX} \equiv L_{\rm IR}/L_{\rm UV}$), and the structural parameters (parametric measurements: the S\'{e}rsic index $n$ and the radius $r_{\rm 20}$; non-parametric measurements: the Gini coefficient $Gini$ and the concentration index $C$). To compare the physical properties of AGNs with those of non-AGNs, we perform Kolmogorov-Smirnov (KS) tests to see whether AGNs have different distributions of these physical properties compared to those of non-AGNs.
The quantity $P_{\rm KS}$ gives the probability that two samples are drawn from the same underlying parent distribution. The critical value of $P_{\rm KS}=0.05$ is the upper limit to verify that AGNs and non-AGNs have different distribution of a physical property at $\geq 2 \sigma$ significance.

\subsection{Stellar Population Properties}
\label{Sect4.1}
No matter a cSFG has a detected AGN or not, the physical parameters of the cSFG can be derived using the CIGALE code.
In this work, we use $M_*$ and SFR estimated from CIGALE. \cite{Shen+20} have confirmed that the difference between the stellar masses estimated by FAST and CIGALE is negligible. The difference dose not change as a function of the stellar mass, even considering the AGN component.
We also estimate the difference between the stellar masses derived from CIGALE and FAST, and the median of the difference is $-0.10^{+0.12}_{-0.33}$, where the 25-75th percentile ranges are nominated.
The difference is small and therefore does not affect the stellar mass distributions of AGNs and non-AGNs and their KS tests shown in this paper.

When the FIR data are included, the CIGALE can better quantify the SFR of host galaxy without being biased by the AGN contamination. This is because most of the FIR emission come from dust absorbing the UV/optical emission. As such, the $\rm sSFR$ can be calculated by the ratio of SFR to stellar mass. The $L_{\rm UV}$ is an estimation of the rest-frame UV luminosity, integrating 1216-3000 \angstrom, and the $L_{\rm IR}$ is the rest-frame 8-1000 $\mu m$ IR luminosity. Both $L_{\rm UV}$ and $L_{\rm IR}$ are in units of $L_{\odot}$.

\begin{figure*}[!t]
  \begin{center}
  \includegraphics[width=0.98\textwidth,angle=0]{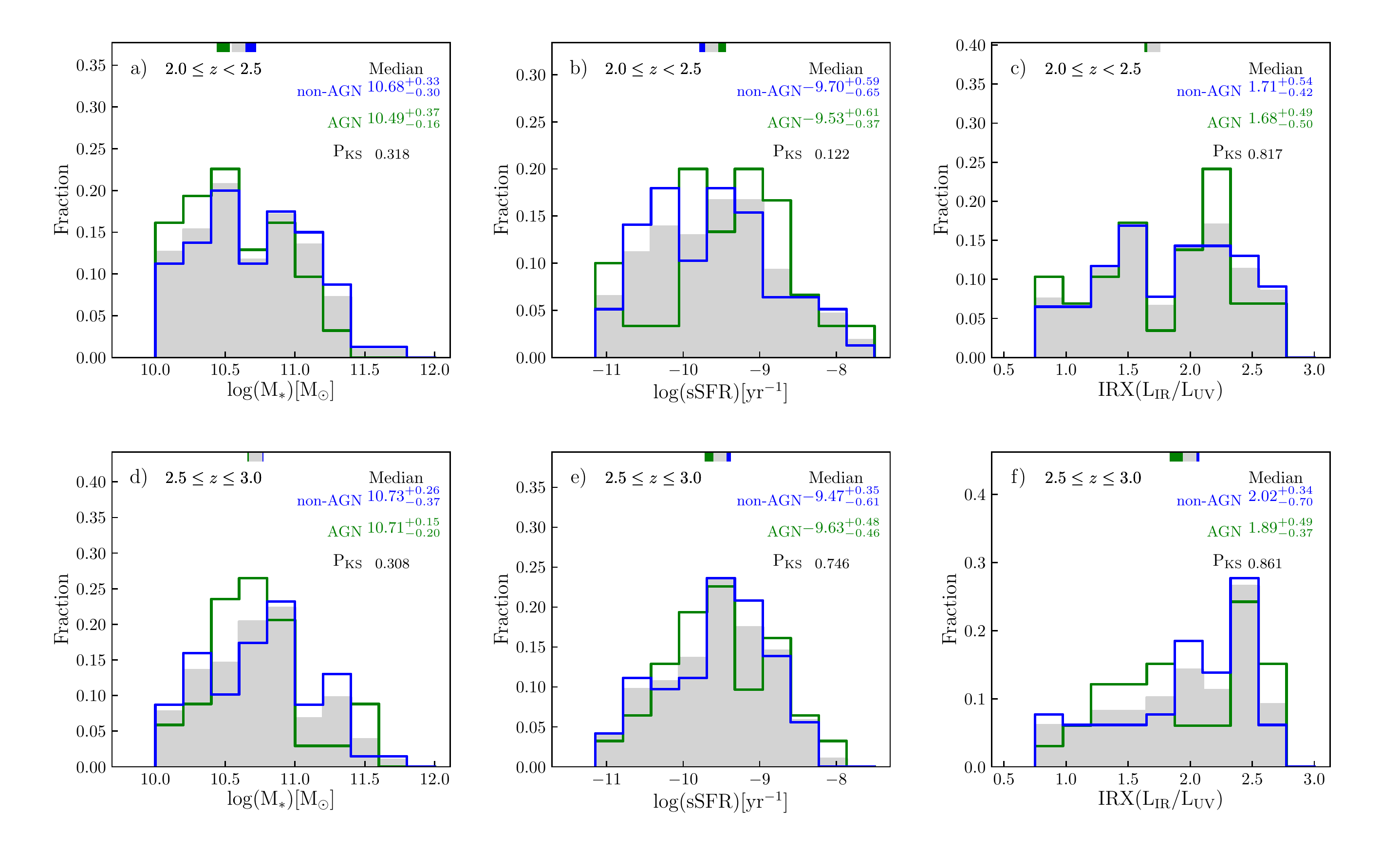}
  \end{center}
  \caption{The distributions of the stellar mass $M_*$, the specific star formation rate $\rm sSFR$ (= ${\rm SFR}/M_\ast$), and the $\rm IRX$ (=$ L_{\rm IR}/L_{\rm UV}$) for AGN (green), non-AGN (blue), and all cSFGs (gray) at $2 <z <2.5$ (top panels) and $2.5 \le z < 3$ (bottom panels). The median values with 25-75th percentile ranges of each distribution for AGNs and non-ANGs are nominated in the corresponding colors. The results of the KS test between AGNs and non-AGNs are marked in black color. The median values are also marked with the corresponding colorful bars at the top x-axis of each panel.}
  \label{fig3}
\end{figure*}

Figure~\ref{fig3} shows the distributions of the physical properties between AGNs and non-AGNs in cSFGs, including the stellar mass $M_*$, the $\rm sSFR$, and the $\rm IRX$. The corresponding medians and 25-75th percentile ranges of each physical property and the KS test probability between AGNs and non-AGNs are also presented in the top right of each panel. In the left panels, the KS test probabilities show that the difference of the stellar mass distributions between AGNs and non-AGNs is negligible for the whole redshift range. The middle panels show their $\rm sSFR$ distributions, for which the KS test probabilities in both redshift bins
are all more than 0.05, which means that given the similar stellar mass distributions, AGNs and non-AGNs have similar $\rm SFR$ distributions.
In order to better reflect the dust-enshrouded level around galactic nucleus, we define a parameter of IRX as the ratio of $L_{\rm IR}$ to $L_{\rm UV}$. If a luminous galaxy is obscured by large amounts of dust and gas, the infrared radiation from dust absorbing UV radiation will be enhanced, resulting in a higher value of IRX. In the right panels of Figure~\ref{fig3}, results of the KS tests show that there is no significant difference of the IRX distributions between AGNs and non-AGNs in the two redshift bins, which means both have similar dust distributions. Moreover, the median values of IRX are getting smaller with cosmic time, which may imply that the gas and dust stored in their host galaxies are gradually being consumed.

\subsection{Parametric Measurements of Structure}
\label{Sect4.2}
\begin{figure}[!t]
\centering
\includegraphics[width=0.6\textwidth]{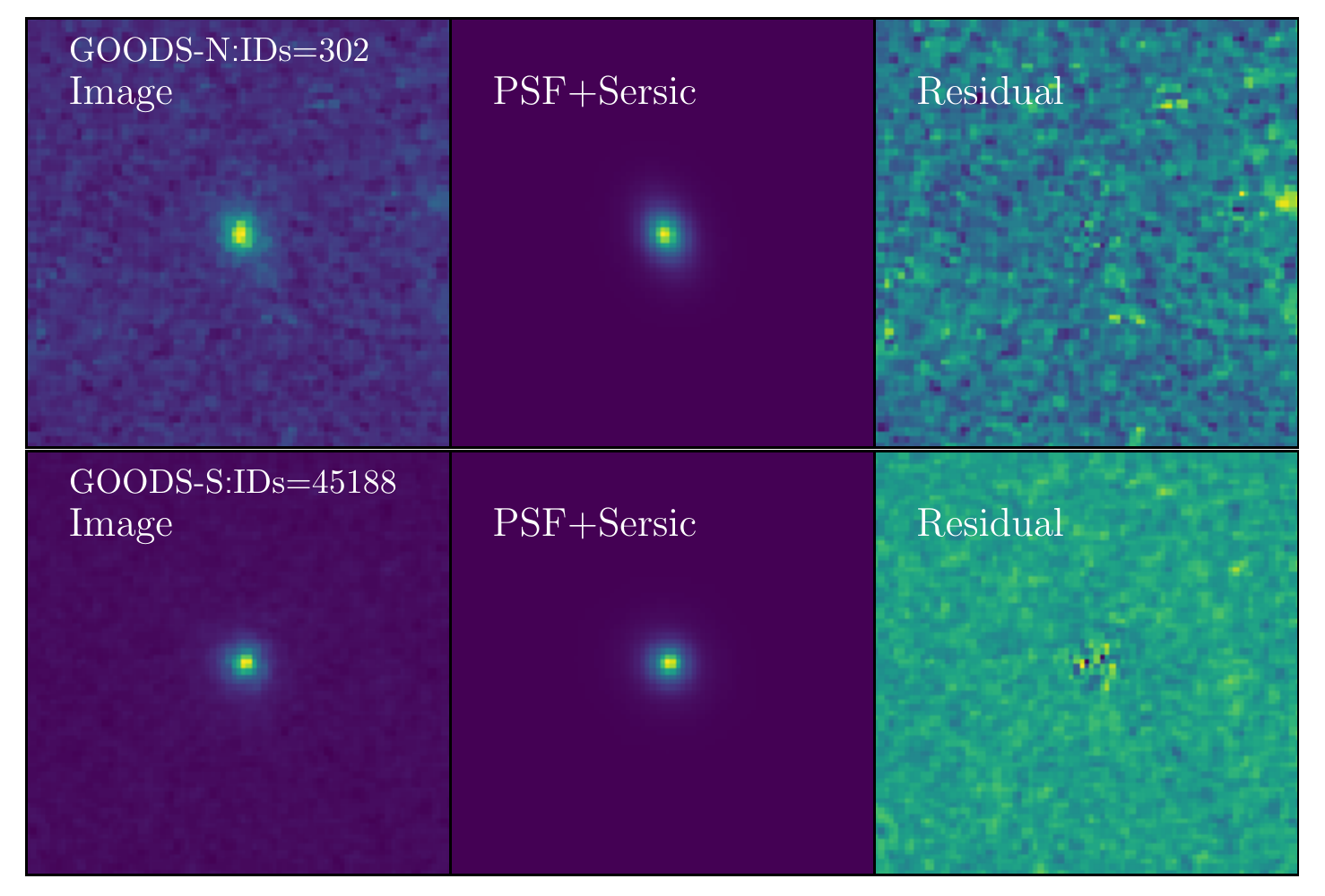}
\caption{Examples of our GALFIT analysis for two of the 37 X-ray-detected AGNs. In the left panels, the \textit{H}-band cutouts are shown with the corresponding field and ID number at the top left of each panel. The corresponding model (PSF+S\'{e}rsic) and residual images are presented in the middle and right panels, respectively. The size of each image is 5\arcsec $\times$ 5\arcsec.}
\label{fig4}
\end{figure}

In this work, the galaxy morphology is traced by the \textit{H}-band image. The galaxy structural parameters, including the S\'{e}rsic index $n$ and the effective radius $r_{\rm e}$, are estimated for each galaxy by \citet{van+der+Wel+12} with a single-component S\'{e}rsic profile in the NIR image using the GALFIT code \citep{Peng+02}. Considering the presence of a bright point-like X-ray-detected AGN in the galaxy center, the measurements on its surface brightness profile may be affected.
Therefore, for the 37 X-ray-detected AGNs, we perform a two-component two-dimensional surface brightness modeling with GALFIT following \citet{Fan+14}, using a point spread function (PSF) model for the nuclear point source and a S\'{e}rsic function for the host galaxy. We constrain the S\'{e}rsic index within a proper range, i.e., $0.1 \leq n\leq 8$.
Being consistent with many previous works showing that the structural measurements of X-ray-detected galaxies should be reliable (\citealt{Kocevski+17,Yang+17, Li+19, Ni+19, Gu+20}), we also find that the central AGN dose not have a significant effect on the effective radius $r_{\rm e}$. Similar to \citet{Barro+13}, we verify again that the cSFGs with X-ray detected AGNs are genuinely compact even after removing the contribution of their central AGN.
Figure~\ref{fig4} shows two examples of our GALFIT analysis for the X-ray-detected AGNs, while for the remaining galaxies, we directly use the structural parameters from \citet{van+der+Wel+12}. Instead of using the derived effective radius $r_{\rm e}$, which contains half of the total light in the best fitting S\'{e}rsic model, we alternatively calculate the more internal radius of a galaxy $r_{\rm 20}$, which should be sensitive to reflect the size change of a compact galaxy. \citet{Miller+19} present that the radius $r_{\rm 20}$, containing 20\% of a galaxy's total luminosity, can be derived with the S\'{e}rsic index $n$ and the effective radius $r_{\rm e}$. And the $r_{\rm 20}$ is closely related to processes controlling the star formation.

\begin{figure*}[!t]
\begin{center}
\includegraphics[width=0.7\textwidth, angle=0]{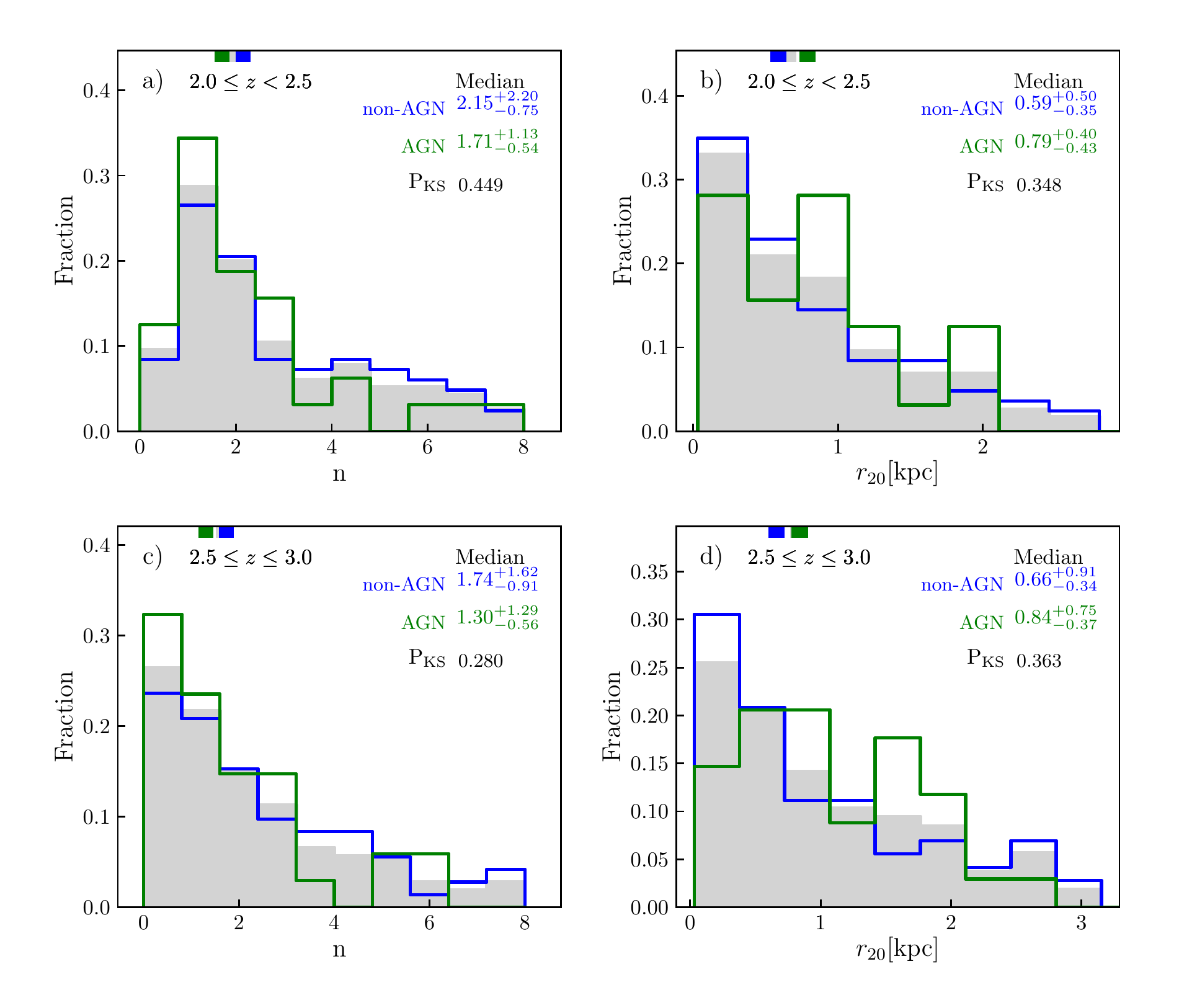}
\end{center}
\caption{Same as Figure~\ref{fig3}, but for the distributions of the S\'{e}sic index $n$ (left panels) and the radius $r_{\rm 20}$ (right panels).}
\label{fig5}
\end{figure*}

Figure~\ref{fig5} shows the distributions of $n$ and $r_{\rm 20}$ for AGNs (green) and non-ANGs (blue) with increasing redshift from top to bottom. The median of the S\'{e}rsic index gets larger with decreasing redshift, which may imply that cSFGs become slightly more compact with cosmic time.
The KS test probabilities for the difference of the $n$ distributions between AGNs and non-AGNs are 0.449 and 0.280 in the lower and higher redshift bins, respectively.
It reconciles with the comparison of the $r_{\rm 20}$ distributions between AGNs and non-AGNs.
At $2<z<3$, the KS test probabilities for the difference of their $r_{\rm 20}$ are not less than 0.05, and the median values of $r_{\rm 20}$ for both AGNs and non-AGNs become smaller with cosmic time. Thus, we conclude that the $n$ and $r_{\rm 20}$ distributions of AGNs are not different from those of non-AGNs.

We confirm that the same conclusion on the difference of the $n$ and $r_{\rm 20}$ distributions between AGNs and non-AGNs would be achieved if we simply consider a AGN sample at $2 < z < 3$, only containing the 29 AGNs selected by MIR+SED or SED methods.

\subsection{Non-parametric Measurements of Structure}
\label{Sect4.3}
To describe the morphological properties of galaxies in our sample, we also perform our own non-parametric structural measurements using the MORPHEUS software. The MORPHEUS has been modified by \citet{Abraham+07} to accommodate new statistics and larger input images. The non-parametric parameters include the Gini coefficient ($Gini$) and the concentration index ($C$).
Since significant bias against results of the non-parametric structural measurements due to the nuclear point source has been found \citep[e.g.,][]{Bohm+13,Pierce+10}, for each of our 37 cSFGs with X-ray-detected AGNs, we subtract the best-fit point source component derived in Section~\ref{Sect4.2} from its \textit{H}-band image and then obtain the image for the underlying host galaxy. While for the remaining galaxies, we directly measure their non-parametric parameters using their original \textit{H}-band images.
In addition, \citet{Lotz+04} and \citet{Lisker+08} have pointed out that the non-parametric structural measurements are strongly dependent on the signal-to-noise ratio per pixel ($S/N_{p.p.}$), especially for $S/N_{p.p.} \leq 2$. Our non-parametric measurements would not suffer from this signal-to-noise ($S/N$) effect because one of our selection criteria, i.e., $\texttt{use\_phot~=~1}$, ensures a reliable detection in $H_{\rm F160W}$ with $S/N_{p.p.}>3$.

Following \citet{Abraham+1994}, to describe the concentration of the galactic surface brightness distribution, the concentration index $C$ is derived by the ratio of the integral flux over $0.3\times$ isophotal radius to that over $1\times$ isophotal radius:
\begin{equation}
\label{eq3£ºC}
   C=\frac{F_{\rm 0.3R}}{F_{\rm R}},
\end{equation}
where $R$ is the isophotal radius for an enclosed size by galaxy isophote at $\sigma$ level above the sky background.
The Gini coefficient $Gini$ is a statistical tool to quantify the unequal light distribution \citep{Lotz+04}.
The $Gini$ is measured by
\begin{equation}
   Gini = \frac{\sum^{N}_{l}(2l-N-1)\mid F_{l}\mid}{\overline{F}N(N-1)},
\label{eq4:Gini}
\end{equation}
where $\overline{F}$ is the mean pixel flux density, $F_{l}$ is the flux density of the $l$-th pixel, and $N$ is the total number of pixels belonging to a galaxy. The $Gini$ describes the relative distribution of the galaxy pixel flux densities.
The $Gini$ is relative to $C$, yet both are not the same. If a galaxy has a large $C$ due to the concentration of light at the center, the $Gini$ may have high probability of being a large value. Conversely, a large $Gini$ dose not correspond to a single bright light at the center (i.e., a large $C$) but may be due to some pixels with large flux densities distributed in the outer region.

\begin{figure*}[!t]
\begin{center}
\includegraphics[width=0.7\textwidth, angle=0]{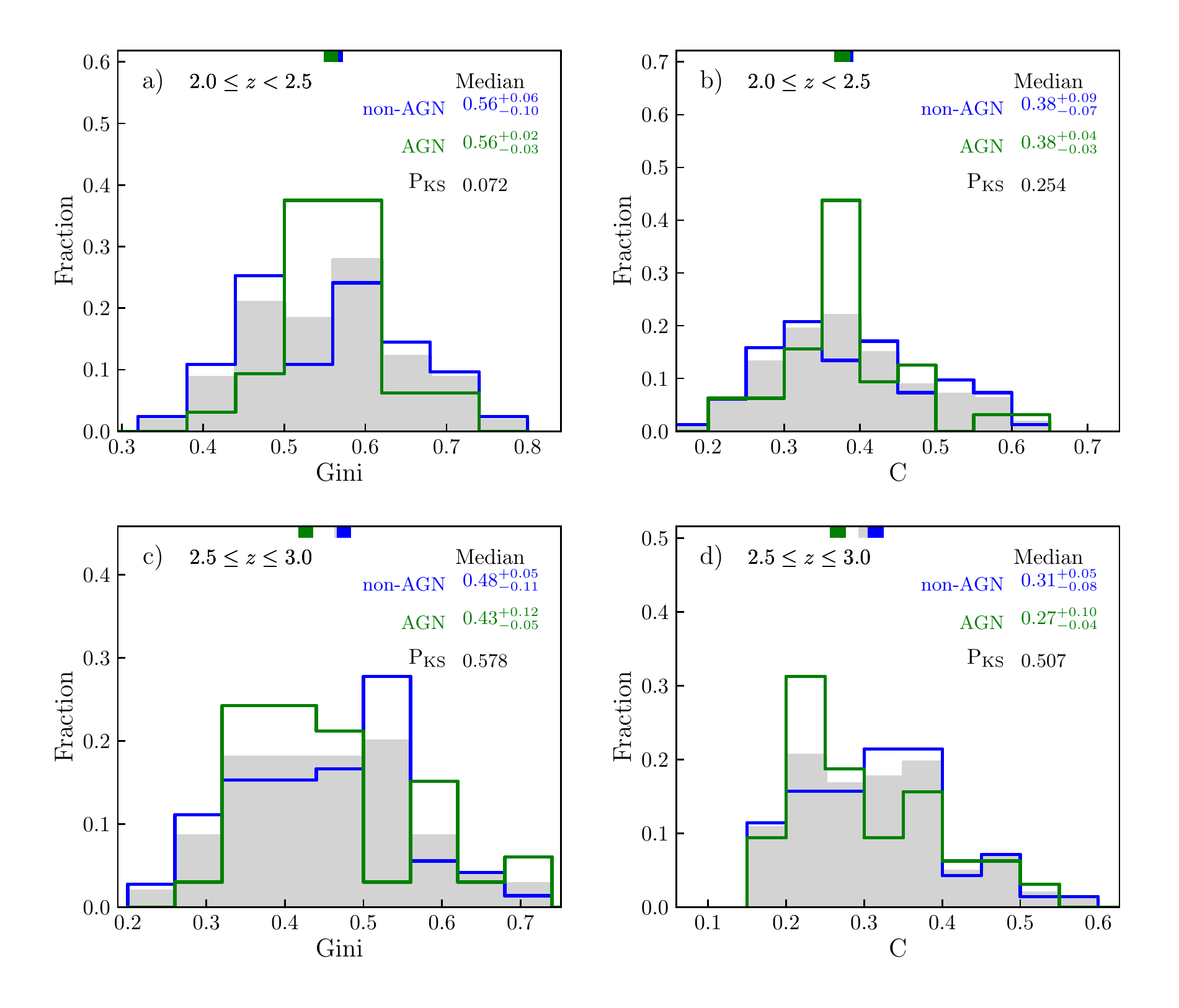}
\end{center}
\caption{
Same as Figure~\ref{fig3}, but for the distributions of the $Gini$ coefficient (left panels) and the concentration index $C$ (right panels).}
\label{fig6}
\end{figure*}

Figure~\ref{fig6} shows the distributions of the non-parametric parameters, $Gini$ and $C$, for AGNs and non-AGNs at $2 < z < 3$.
The KS test probabilities for the difference of the $Gini$ and $C$ distributions between AGNs and non-AGNs are more than 0.05 at $2<z<3$, which implies that both $Gini$ and $C$ of the AGN hosts are drawn from the same distributions as those of non-AGNs. The median values of $Gini$ and $C$ for both AGNs and non-AGNs increase \textbf{slightly} with decreasing redshift.

\begin{figure*}[!t]
\begin{center}
\includegraphics[width=0.75\textwidth,angle=0]{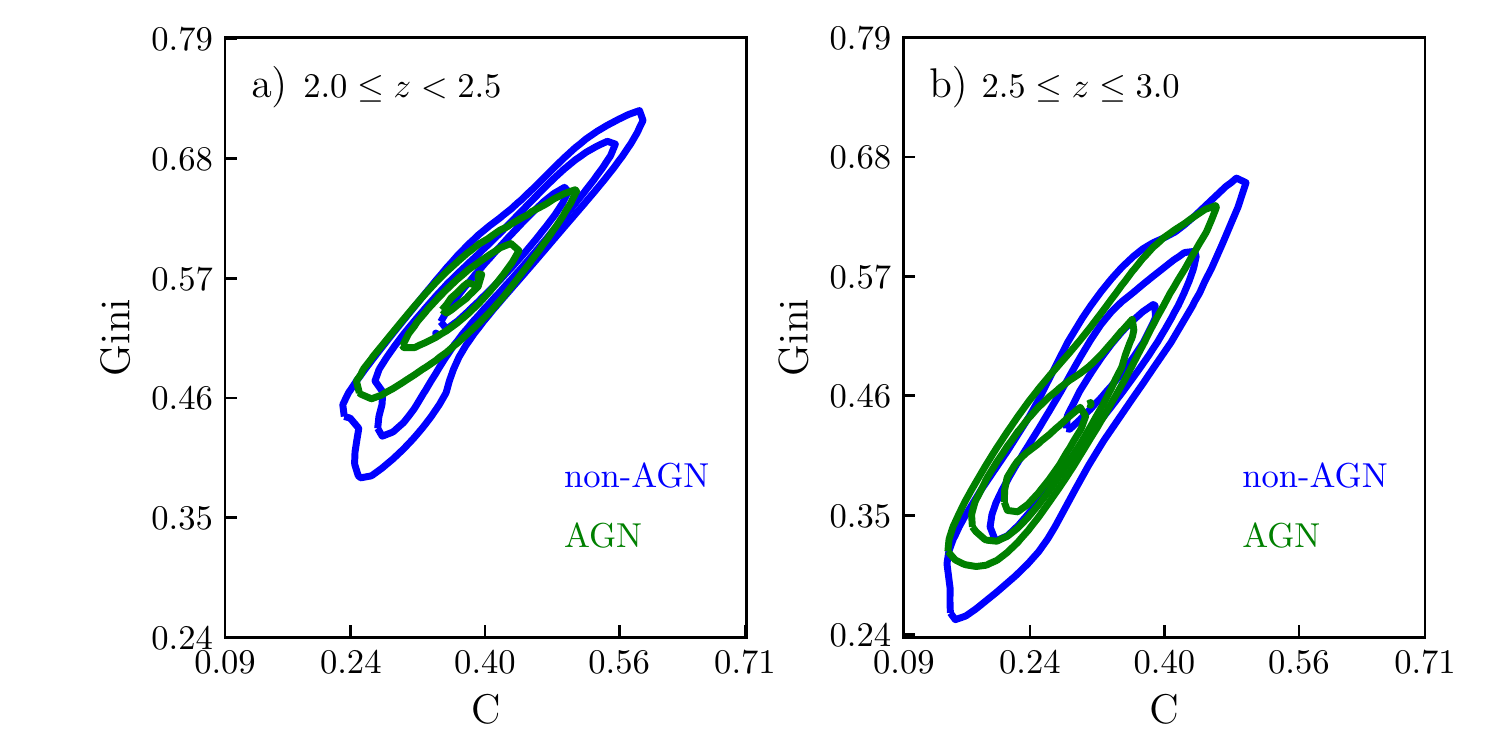}
\end{center}
\caption{
The correlations between the $Gini$ coefficient and the concentration index $C$ for AGNs (green) and non-AGNs (blue) in two redshift bins.
The contour levels for AGNs (green) and non-AGNs (blue) trace the 20\%, 50\%, and 80\% of grid counts with their relative densities in the $Gini$-$C$ relation which are sorted in descending order.}
\label{fig7}
\end{figure*}

For a more intuitive comparison, the contour maps of $Gini$ versus $C$ for AGNs (in green lines) and non-AGNs (in blue lines) in the two redshift intervals are displayed in Figure~\ref{fig7}.
The contours of both AGNs and non-AGNs are similar in each redshift bin. And
both AGNs and non-AGNs occupy higher $Gini$-$C$ space with decreasing redshift, which might indicates that cSFGs seem to be slightly more concentrated with cosmic time.

Again, the same conclusion on the difference of the $Gini$ and $C$ distributions between AGNs and non-AGNs would be achieved if we simply consider a AGN sample at $2 < z < 3$, only containing the 29 AGNs selected by MIR+SED or SED methods.

\section{Environmental Effect}
\label{Sect5}
Environment plays a crucial role in galaxy evolution (e.g., \citealt{Muldrew+12, Darvish+15}), especially at lower redshift (e.g., \citealt{Ilbert+13}).
Different processes can trigger the compaction event, such as major or minor merger, violent disk instabilities, counter-rotating streams, tidal interactions and perturbation owing to giant clumps (e.g., \citealt{Hopkins+06,Dekel+Burkert+14,Zolotov+15}). And galaxies residing in overdensities may easily merge and interact with nearby galaxies, then subsequently causing centrally concentrated starbursts or AGN feedback (e.g., \citealt{Belli+17,Maltby+18,Belli+19}).
To analyse the environmental effects on AGNs and non-AGNs in cSFGs, we improve the traditional measurement of environment using a Bayesian metric (Gu et al. 2020, in preparation).

The traditional indicator of local environment depends on the \textit{N}th nearest neighbour or the count of neighboring galaxies within a fixed aperture (e.g., \citealt{Dressler+1980}). Alternatively, the modified indicator of environment uses the Bayesian metric to take the distances of all \textit{N} nearest neighbors into consideration \citep{Ivezic+05,Cowan+Ivezic+08}. We firstly build a magnitude-limited sample at $z=2$-$3$ with $H_{F160W}<25.5$ for the measurement of environment.
The magnitude cut can make sure the uncertainty of photometric redshift $\Delta z/ (1+z) \approx 0.02$ and most high-redshift faint galaxies detected \citep{Skelton+14}.
The local surface density of each galaxy is estimated by $\Sigma'_{N}=1/\Sigma_{i=1}^{N}d^2_{i}$, where $d_i$ is the projected distance to \textit{i}th nearest neighbor within a redshift slice ($|\Delta z| < \sigma_{\rm z}(1+z)$, $\sigma_{\rm z}=0.02$).
Then the dimensionless overdensity $1+\delta'_{N}$, describing the relative density of environment, is estimated by
\begin{equation}\label{eq5_overdensity}
  1+\delta'_{N} = \frac{\Sigma'_{N}}{\langle \Sigma'_{N}\rangle}=\frac{\Sigma'_{N}}{k'_{N}\Sigma_{\rm surface}},
\end{equation}
where $\Sigma_{\rm surface}$ is the surface number density within a given redshift slice. The correction factor $k'_{N}$ is used to describe the intrinsic correlation between $\Sigma_{\rm surface}$ and $<\Sigma'_{N}>$. Due to the compaction event possibly triggered by the merger of neighboring galaxies, we adopt $N=3$ and $k'_{3}= 0.08$ in this work, where the adopted number of neighboring galaxies $N$ does not affect our main results.
The overdensity $1+\delta'_{N} =1$ (i.e., $\log (1+\delta'_{N}) =0$) represents a standard level of average density, while the overdensity $1+\delta'_{N}$ over and below this standard level indicates the excess and lack of the environmental density.

\begin{figure*}[!t]
\begin{center}
\includegraphics[width=0.65\textwidth, angle=0]{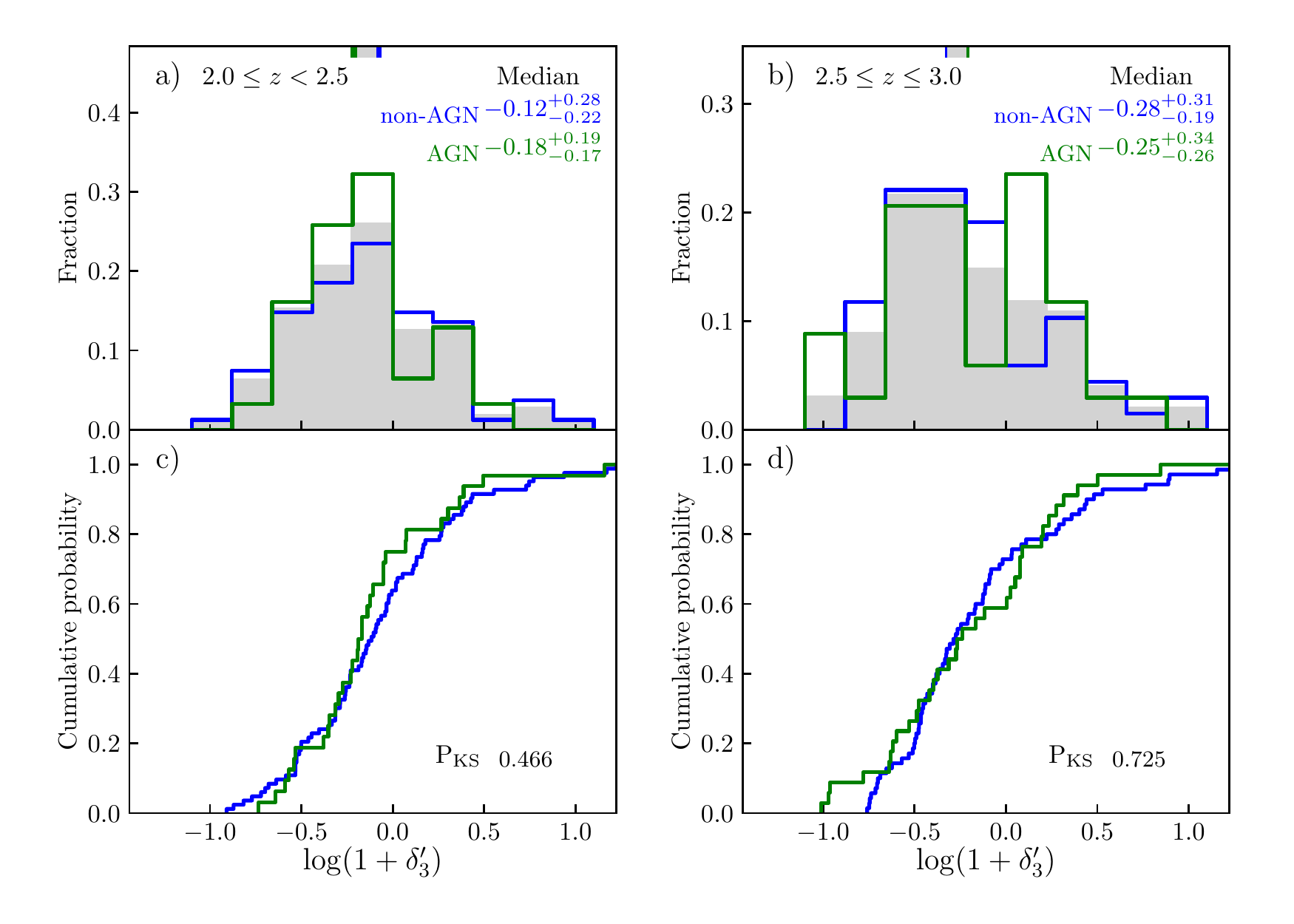}
\end{center}
\caption{Distributions and cumulative probabilities of the overdensity for AGNs (green) and non-AGNs (blue) at $2< z < 3$. The distribution of all cSFGs is shown in the gray color. Their corresponding median values are given in panels of the first row, and the probabilities of KS tests are shown in panels of the second row.}
\label{fig8}
\end{figure*}

The comparison of the environment between AGNs and non-AGNs is shown in Figure~\ref{fig8}. The distributions of overdensities for AGNs and non-AGNs in our cSFG sample are shown in the upper panels, while in the bottom panels, the corresponding cumulative distribution functions are present. The KS test probabilities indicate that the environment around cSFG with AGNs is similar to that around those with non-AGNs. Cosmological hydrodynamical simulations of galaxy formation suggest that the formation of compact star-forming systems is caused by dissipative contraction \citep{Dekel+Burkert+14}, such as major merger or violent disk instabilities. After the dissipative compaction, both AGNs and non-AGNs in cSFGs reside in a similar environment. Even in the large scale environment, \citet{Krishnan+20} still prove that the AGN does not reside in a ``special'' environment. It implies that the AGN activity is potentially triggered through internal secular processes, such as gravitational instabilities and/or dynamical friction. \citet{Bournaud+11} and \citet{Chang+17} also point out that the AGN activity can be triggered by the inflow of gas that has been tidally stripped from the companion during the process of compaction.

\section{Discussion}
\label{Sect6}
The cSFGs are regarded as the progenitors of cQGs in previous works (e.g., \citealt{Barro+13,Barro+14,Rangel+14,Fang+15,Lu+19,Gu+20}). The cSFGs formed by gas-rich processes are found to possess a high fraction of X-ray-selected AGN (e.g, \citealt{Barro+14,Fang+15,van+Dokkum+15,Gu+20}) and an average quenching timescale $t_{\rm q}< 1$ Gyr (e.g., \citealt{Barro+14,van+Dokkum+15,Lu+19,Gu+19}). It implies that AGN feedback may have played important roles in transforming cSFGs to cQGs, which motivates us to analyse the difference of physical properties for AGNs and non-AGNs in cSFGs.

In this work, we analyse the difference of several physical properties of cSFG with AGNs and non-AGNs. By comparing the sSFR distribution of AGNs to that of non-AGNs, it is found that there is no obvious evidence that AGNs can quench or promote the star formation of their host galaxies (\citealt{Hatziminaoglou+10, Harrison+12, Stanley+15, Xu+15}). The results of the structural analysis manifest that AGNs have structure similar to non-AGNs, which is consistent with the previous findings (\citealt{Kocevski+12,Fan+14,Kocevski+17}). It may indicate that major mergers could induce the formation of cSFGs as suggested by simulations, but may not be necessary for triggering AGN activities (\citealt{Fan+14}). The results of both AGN and non-AGN living in similar environments also hint that internal secular processes should play a crucial role in triggering AGN activities.

Considering the redshift evolution of structures, it can be found that the compact star-forming systems formed by dissipative contraction seem to get slightly concentrated with cosmic time due to the gradual consumption of dust and gas. As shown in Figure~\ref{fig3}, the IRX values, a proxy for dust obscuration, tend to be smaller with cosmic time. In Figure~\ref{fig7}, the contours of AGNs and non-AGNs shift to the higher $Gini$ versus $C$ space with decreasing redshift. Even if we keep a same average level of the $S/N$ of cSFGs between in the higher and lower redshift interval, the redshift evolution of the contours of $Gini$ and $C$ still seem to exist. Similar to the results from \citet{Gu+20}, the cSFGs have a tendency to become slightly more compact with cosmic time, accompanying the consumption of available gas and dust. This is consistent with the results from \citet{Chang+17}, where the AGN host SFGs undergone a process of dynamical contraction.

\section{Summary}
\label{Sect7}
In this work, we have constructed a sample of 221 massive cSFGs with $\log_{10}(M_{*}/M_{\odot}) \geq 10$ at $2<z<3$ in five 3D-HST/CANDELS fields, in which 66 AGNs are selected by the X-ray, the MIR criteria, and/or the SED fitting. We present analyses of the differences in several physical properties of cSFGs with AGNs and non-AGNs, such as the stellar mass, the $\rm sSFR$, the $\rm IRX$, the structural parameters (i.e., $n$, $r_{\rm 20}$, $Gini$, and $C$), and the environment. Our main conclusions are summarized as follows :

\begin{enumerate}
  \item We integrate multiple AGN selection methods to constitute the most complete census of AGNs. The AGN fraction in cSFGs is $\sim 30 \pm 3.1\%$, which is higher than the one obtained using a single selection method and is statistically closer to the actual truth.
  \item By comparing physical properties of cSFGs with AGNs and non-AGNs, it is found that AGNs in cSFGs have similar stellar mass ($M_*$), specific star formation rate (sSFR), ratio of $L_{\rm IR}$ to $L_{\rm UV}$ (IRX $\equiv$ $L_{\rm IR}/L_{\rm UV}$), S\'{e}rsic index ($n$), internal radius ($r_{\rm 20}$), Gini coefficient ($Gini$), and concentration index ($C$) to non-AGNs within the same redshift bin of $2 < z < 2.5$ and $2.5 \le z < 3$.
  \item After dissipative compaction events, such as major merger or violent disk instabilities, the compact star-forming systems are formed, then their structures might become slightly more compact with cosmic time due to the gradual consumption of dust and gas. During the redshift evolution of cSFGs, both AGNs and non-AGNs reside in a similar environment, and the AGN activities may be triggered by some  internal secular processes.

\end{enumerate}

\acknowledgments
This work is based on observations taken by the 3D-HST Treasury Program (GO 12177 and 12328) with the NASA/ESA \textit{HST}, which is operated by the Association of Universities for Research in Astronomy, Inc., under NASA contract NAS5-26555.
This work is supported by the National Science Foundation of China (grant Nos. 11673004, 11873032, and 11433005) and the Research Fund for the Doctoral Program of Higher Education of China (20133207110006). G.W.F. acknowledges support from Yunnan young and middle-aged academic and technical leaders reserve talent program (201905C160039), Yunnan ten thousand talent program - young top-notch talent and Yunnan Applied Basic Research Projects (2019FB007).

\bibliography{reference}{}
\bibliographystyle{aasjournal}

\end{document}